\documentclass[twocolumn]{aastex63}
\usepackage{times}
\usepackage{amsmath}
\usepackage{textcomp}
\usepackage{amssymb}
\usepackage{natbib}
\usepackage{float}
\usepackage{color}
\usepackage{graphicx}
\usepackage{epstopdf}
\newcommand{\Msun}{\ensuremath{M_{\odot}}}
\newcommand{\lum}{erg\,s$^{-1}$}
\newcommand{\fermi}{{\it Fermi}}

\newcommand{\gm}{$\gamma$}
\received{\today}
\revised{\today}
\accepted{\today}
\submitjournal{ApJL}

\shorttitle{The First $z>3$ BL Lac Object}
\shortauthors{Paliya et al.}

\begin{document}

\title{The First Gamma-ray Emitting BL Lacertae Object at the Cosmic Dawn}

\correspondingauthor{Vaidehi S. Paliya}
\email{vaidehi.s.paliya@gmail.com}

\author[0000-0001-7774-5308]{Vaidehi S. Paliya}
\affiliation{Deutsches Elektronen Synchrotron DESY, Platanenallee 6, 15738 Zeuthen, Germany}

\author[0000-0002-3433-4610]{A. Dom{\'{\i}}nguez}
\affiliation{Institute of Particle and Cosmos Physics (IPARCOS), Universidad Complutense de Madrid, E-28040 Madrid, Spain}
\affiliation{Department of EMFTEL, Universidad Complutense de Madrid, E-28040 Madrid, Spain}

\author[0000-0003-4187-7055]{C. Cabello}
\affiliation{Institute of Particle and Cosmos Physics (IPARCOS), Universidad Complutense de Madrid, E-28040 Madrid, Spain}
\affiliation{Department of Physics of the Earth and Astrophysics, Universidad Complutense de Madrid, E-28040 Madrid, Spain}

\author[0000-0002-9334-2979]{N. Cardiel}
\affiliation{Institute of Particle and Cosmos Physics (IPARCOS), Universidad Complutense de Madrid, E-28040 Madrid, Spain}
\affiliation{Department of Physics of the Earth and Astrophysics, Universidad Complutense de Madrid, E-28040 Madrid, Spain}

\author[0000-0003-1439-7697]{J. Gallego}
\affiliation{Institute of Particle and Cosmos Physics (IPARCOS), Universidad Complutense de Madrid, E-28040 Madrid, Spain}
\affiliation{Department of Physics of the Earth and Astrophysics, Universidad Complutense de Madrid, E-28040 Madrid, Spain}

\author[0000-0002-4935-9511]{Brian Siana}
\affiliation{Department of Physics and Astronomy, University of California Riverside, Riverside, CA 92521, USA}

\author[0000-0002-6584-1703]{M. Ajello}
\affiliation{Department of Physics and Astronomy, Clemson University, Kinard Lab of Physics, Clemson, SC 29634-0978, USA}

\author[0000-0002-8028-0991]{D. Hartmann}
\affiliation{Department of Physics and Astronomy, Clemson University, Kinard Lab of Physics, Clemson, SC 29634-0978, USA}

\author[0000-0001-6150-2854]{A. Gil de Paz}
\affiliation{Institute of Particle and Cosmos Physics (IPARCOS), Universidad Complutense de Madrid, E-28040 Madrid, Spain}
\affiliation{Department of Physics of the Earth and Astrophysics, Universidad Complutense de Madrid, E-28040 Madrid, Spain}

\author[0000-0002-4998-1861]{C. S. Stalin}
\affiliation{Indian Institute of Astrophysics, Block II, Koramangala,Bengaluru, Karnataka 560034, India}

\begin{abstract}
One of the major challenges in studying the cosmic evolution of relativistic jets is the identification of the high-redshift ($z>3$) BL Lacertae objects, a class of jetted active galactic nuclei characterized by their quasi-featureless optical spectra. Here we report the identification of the first \gm-ray emitting BL Lac object, 4FGL~J1219.0+3653 (J1219), beyond $z=3$, i.e., within the first two billion years of the age of the Universe. The optical and near-infrared spectra of J1219 taken from 10.4 m Gran Telescopio Canarias exhibit no emission lines down to an equivalent width of $\sim$3.5 \AA~supporting its BL Lac nature. The detection of a strong Lyman-$\alpha$ break at $\sim$5570 \AA, on the other hand, confirms that J2119 is indeed a high-redshift ($z\sim3.59$) quasar. Based on the prediction of a recent BL Lac evolution model, J1219 is one of the only two such objects expected to be present within the comoving volume at $z=3.5$. Future identifications of more $z>3$ \gm-ray emitting BL Lac sources, therefore, will be crucial to verify the theories of their cosmic evolution.
\end{abstract}


\section{Introduction}{\label{sec:Intro}}
Bl Lacertae objects (BL Lacs) are a sub-class of radio-loud active galactic nuclei (AGN) hosting closely aligned relativistic jets, also known as blazars. Their characterization is based on the observation of a quasi-featureless optical spectrum with no or weak emission lines with rest-frame equivalent width (EW) $<$5\AA~\citep[e.g.,][]{1991ApJ...374..431S}. It has been predicted that the dominance of the Doppler boosted synchrotron radiation originated from the jet leads to the non-thermal continuum swamping the emission lines. However, the optical spectra of many nearby ($z<0.8$) BL Lacs exhibit prominent stellar absorption features even though no emission lines are seen \citep[e.g.,][]{2011MNRAS.413..805P}. Therefore, the lack of strong emission lines may be directly connected to an inefficient accretion process, which is unable to photo-ionize the broad line region (BLR) clouds \citep[][]{2011MNRAS.414.2674G}.

The foremost challenge offered by the featureless optical spectra of BL Lacs is to reliably determine their distances. The redshift determination is fundamental to study the radiative processes and the cosmic evolution of relativistic jets. For example, the prediction about the lack of luminous, high-frequency peaked BL Lacs \citep[][]{1998MNRAS.299..433F} has been challenged by arguing that it could be due to selection effects \citep[e.g.,][]{2012MNRAS.420.2899G}, since any blazar sample is dominated by the broad-line (EW$>$5\AA) blazars or so-called flat spectrum radio quasars (FSRQs). The powerful, high-redshift ($z>2$) BL Lacs, even if they exist, may remain unknown due to their lineless optical spectra. These objects are also key to explore the evolution of the extragalactic background light at high redshifts \citep[e.g.~][]{2011MNRAS.410.2556D, 2015ApJ...813L..34D, 2018Sci...362.1031F}. For all these reasons, various attempts have been made to identify high-redshift BL Lacs by determining their spectroscopic/photometric redshifts or upper/lower limits to them \citep[see, e.g.,][]{2012A&A...538A..26R,2017ApJ...844..120P,2016AJ....151...35L,2018ApJ...861..130L}.

The fourth catalog of AGN detected with the Large Area Telescope (LAT) on board {\it Fermi Gamma-ray Space Telescope} \citep[4LAC;][]{2020ApJ...892..105A} contains only three spectroscopically confirmed $z>2$ BL Lacs, with the most distant at $z=2.471$. For a comparison, there are 87 FSRQs above $z=2$ in 4LAC with the farthest being at $z=4.31$ \citep[see also][]{2017ApJ...837L...5A}. The \gm-ray detection is crucial as it unambiguously confirms the presence of a beamed relativistic jet and hence the blazar nature of a radio-loud quasar. Among the 10 most distant ($z\geq3$) ones, 9 are classified as FSRQs and one object, 4FGL J1219.0+3653 (hereafter J1219), associated with the radio source NVSS~J121915+365718, is flagged as a blazar of uncertain type (BCU\footnote{Blazar candidates of uncertain type or BCUs are \gm-ray emitting sources whose multi-wavelength properties are similar to blazars, e.g., a flat radio spectrum, but lack spectral characterization needed to classify them as a BL Lac or FSRQ \citep[][]{2015ApJ...810...14A}.}). The only available Sloan Digital Sky Survey (SDSS) spectrum of J1219 exhibits the Lyman-$\alpha$ cut-off at $\sim$5570 \AA, thus supporting its high-redshift nature. More importantly, it does not show any broad emission lines, thus hinting it to be a candidate BL Lac object.  

The identification of a \gm-ray emitting BL Lac beyond $z=3$ can have major implications on the cosmic evolution of relativistic jets. This is because the presence of a single blazar indicates the existence of hundreds of misaligned jetted systems with similar intrinsic properties in the same cosmic volume. Therefore, we carried out deep optical and near-infrared (NIR) spectroscopy of J1219 with the 10.4 m Gran Telescopio Canarias (GTC) to characterize the physical nature of the source. In this letter, we present the derived results and report the spectroscopic identification of the first BL Lac found within the first two billion years of the universe's age. Throughout, we adopt the following cosmology parameters: $H_0=67.8$~km~s$^{-1}$~Mpc$^{-1}$, $\Omega_m = 0.308$, and $\Omega_\Lambda = 0.692$ \citep[][]{2016A&A...594A..13P}

\section{OSIRIS and EMIR Data Reduction}\label{sec:analysis}
In order to cover the target rest-frame for both the Lyman-$\alpha$ and the optical regions, where potential emission lines could be found, long-slit spectroscopic observations with two different instruments have been employed:
OSIRIS \citep[Optical System for Imaging and low-Intermediate-Resolution Integrated
Spectroscopy;][]{2013hsa7.conf..868C}, and EMIR \citep[Espectr\'{o}grafo Multiobjeto
Infra-Rojo;][]{2019hsax.conf..526G}. These observations were carried out
under two Director Discretionary Time (DDT) proposals, which were promptly approved
and completed.

\begin{figure}[t!]
\hbox{
\includegraphics[scale=0.17]{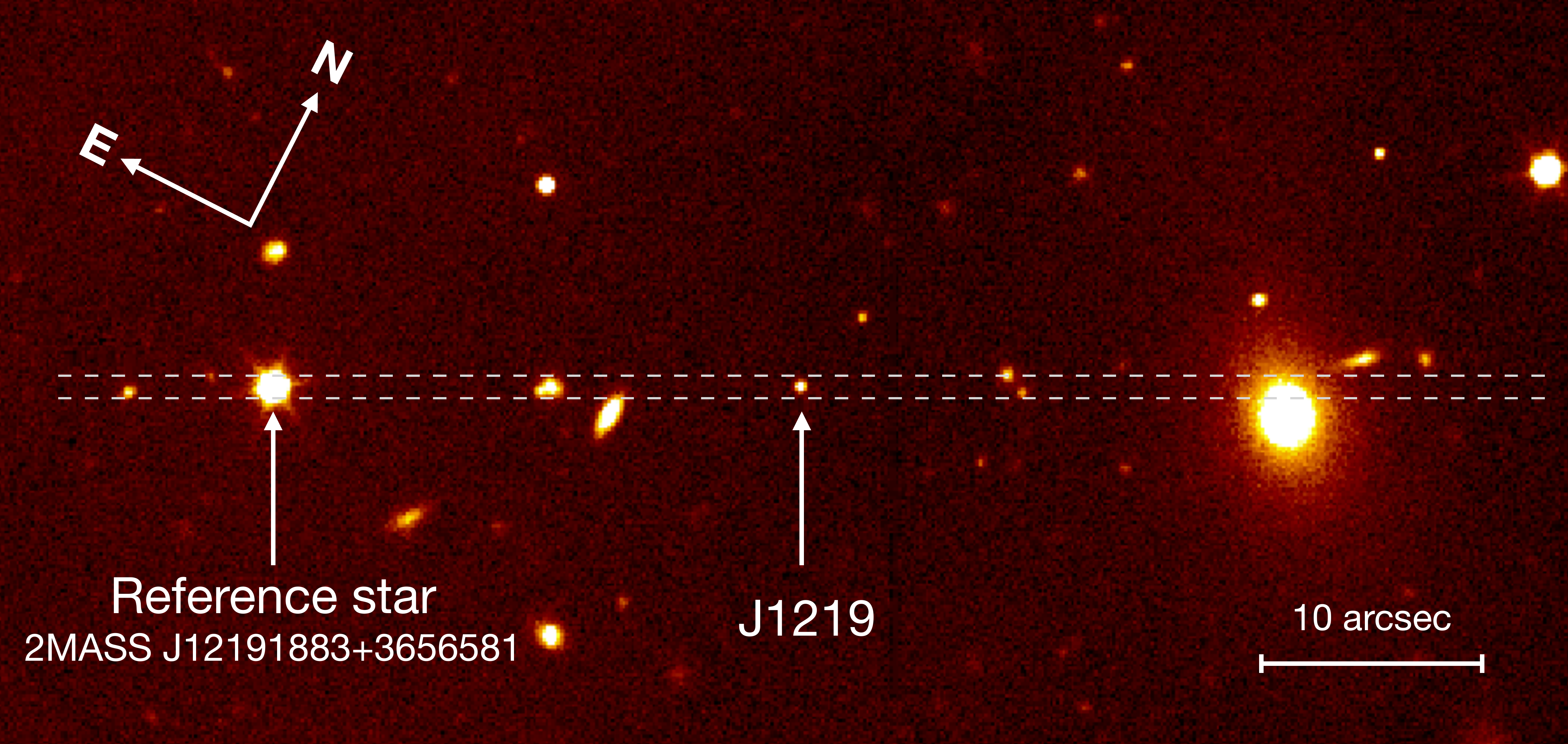} 
}
\caption{Detail of the acquisition image for J1219 obtained with OSIRIS, using the Sloan~r' filter. The dashed lines illustrate an 1~arcsec-width long slit placed over the reference star and the target, the same slit width employed with both OSIRIS and EMIR.}
\label{fig:acquisition_J1219}
\end{figure}

A bright (G=16.7~mag) star, 2MASS~J12191883+3656581, was used during the acquisition procedure, placing the star within the corresponding 1~arcsec-width slits and using a position angle of $116\fdg5$ (Figure~\ref{fig:acquisition_J1219}). This turned out to be important for the NIR observations with EMIR, where the quasar was not visible through the acquisition images obtained after the long-slit configuration of the instrument Cold Slit Unit. 

Considering the faintness of the target, a careful data reduction process was carried out. A full description will be published elsewhere (C. Cabello, PhD thesis, in preparation), and we briefly outline the adopted steps below.

\subsection{OSIRIS spectroscopic data}
\begin{figure*}[t!]
\hbox{
\includegraphics[scale=0.36]{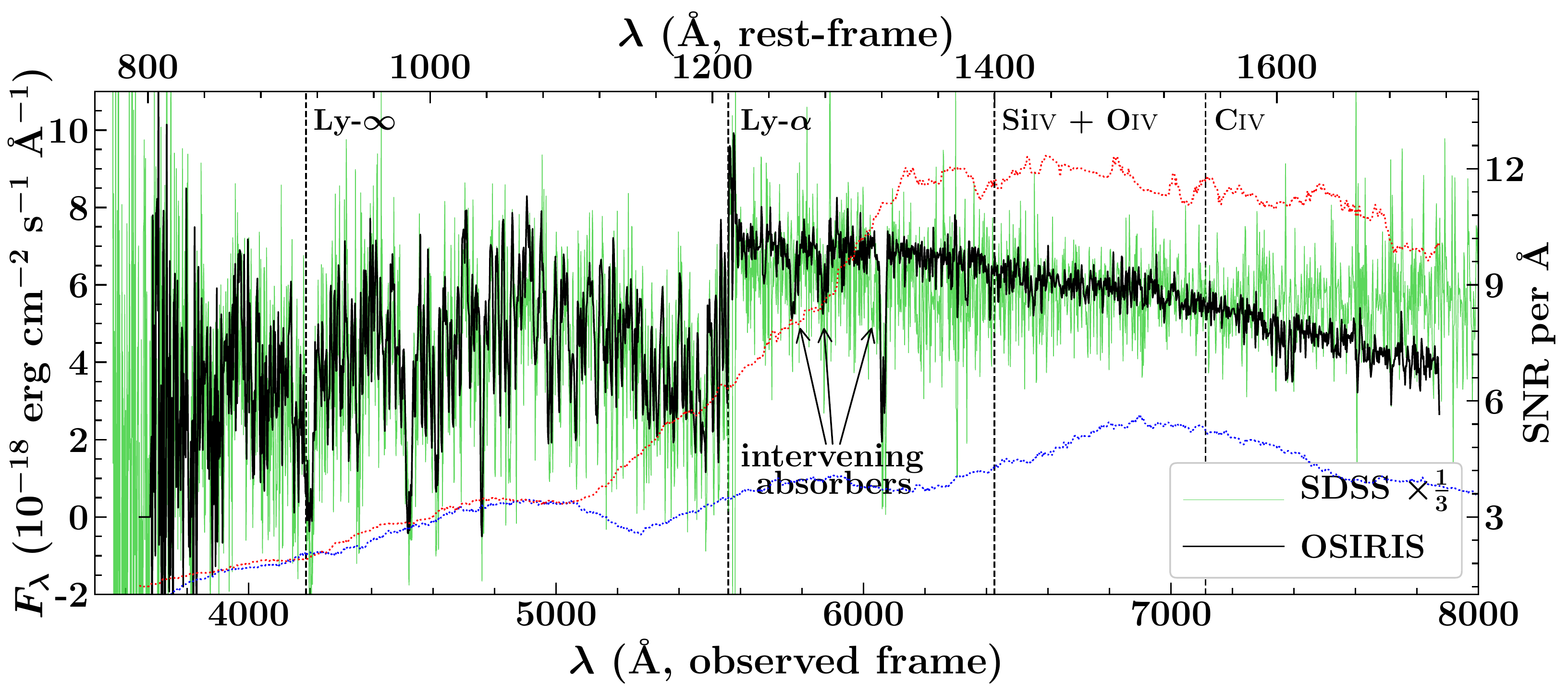} 
\includegraphics[scale=0.29]{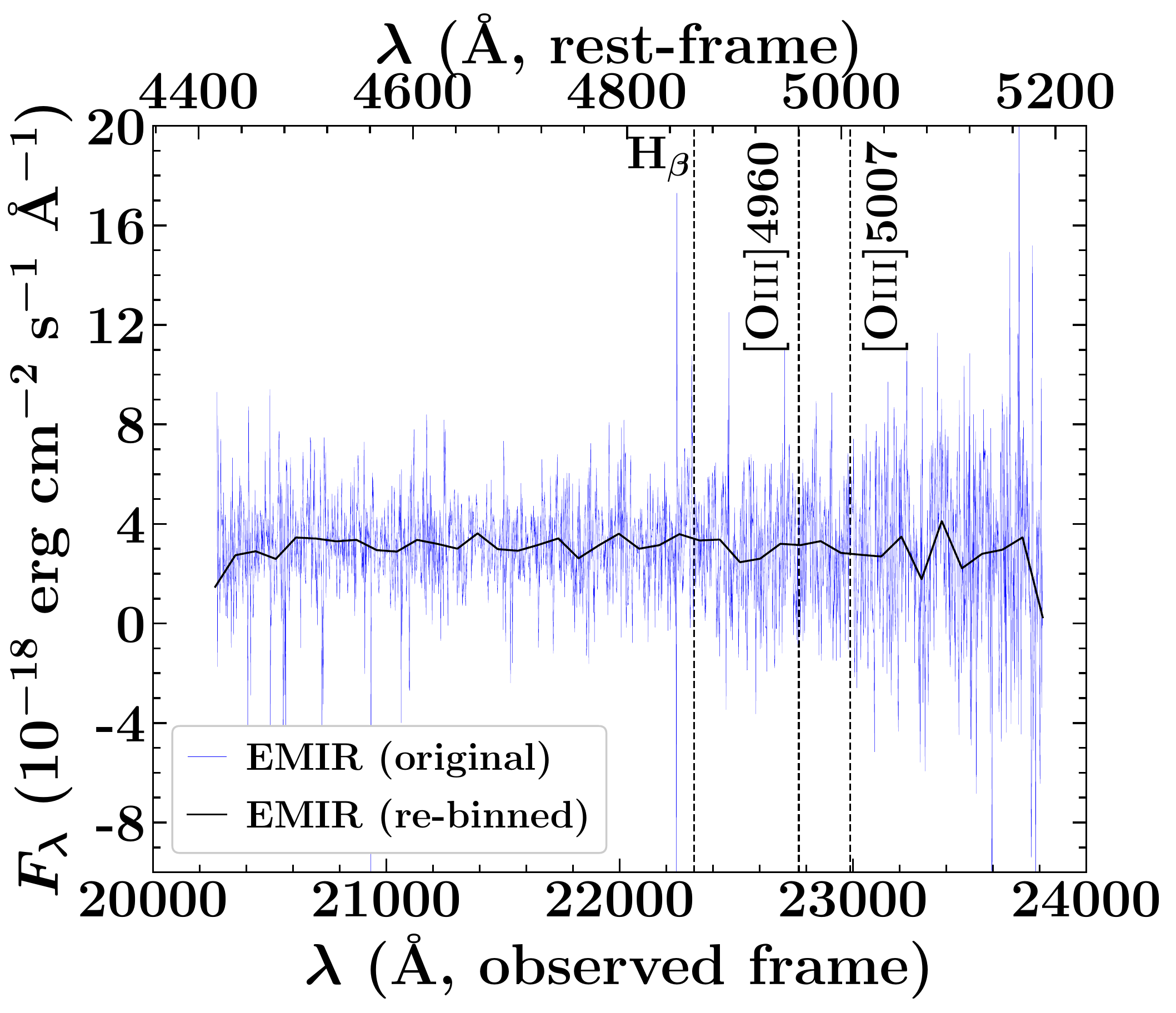}
}
\hbox{\hspace{2.1cm}
\includegraphics[scale=0.5]{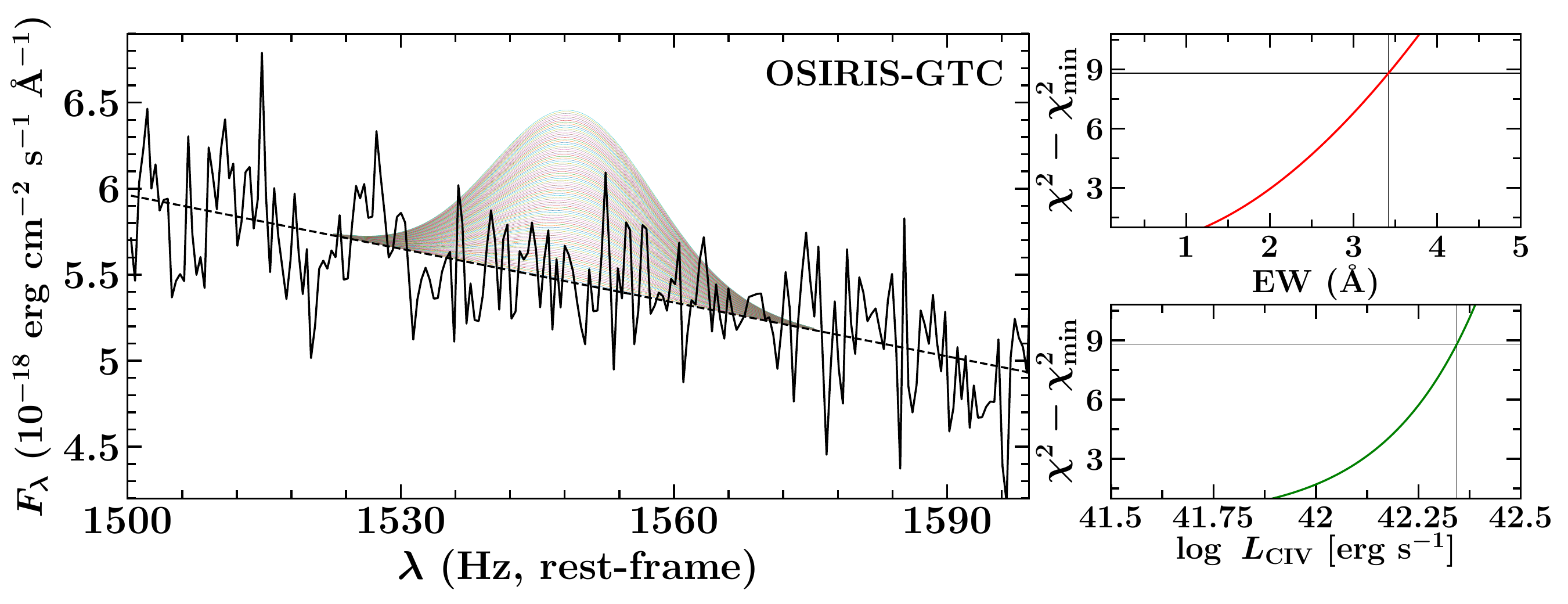}
}
\caption{Top: the optical and NIR spectra of J1219 taken with OSIRIS (left) and EMIR (right). We also show the SDSS spectrum with green lines, after dividing it by a factor of 3 for a comparison. The expected locations of a few prominent emission lines are labeled with vertical dashed lines. The red and blue dotted lines represent the wavelength dependent SNR for the OSIRIS and SDSS spectra, respectively. In the EMIR plot, we show the data without binning (light blue) and with a binning of 50 pixel (black) needed to achieve a SNR of $\sim$5 at the center of the spectrum. Bottom: The rest-frame OSIRIS spectrum when fitted with a power-law (black dashed line) and a single Gaussian function with a variable EW or C~{\sc iv} line luminosity. In the right panels, we plot the variation of the derived $\chi^2$ as a function of the considered parameters. The vertical lines highlight the upper limits beyond which $\chi^2 > \chi^2$ (99.7\%).}
\label{fig:osiris_emir}
\end{figure*}
The observations with OSIRIS were performed on 2020 May 26 (seeing~$\sim1.3$~arcsec), using the R1000B grism, which provides an intermediate spectral resolution \mbox{$R\sim1018$}, with a \mbox{2.06~\AA/pixel} sampling, and a spatial scale of 0.508~arcsec/pixel (with a $2\times2$~binning). The total exposure on-source was 7200 sec divided in three individual exposures of 1200 sec each. Bias correction, trimming, rotation, and flat-fielding were performed using Python scripts. Cosmic rays were removed using
the package \texttt{cleanest} \citep[][]{2020ASPC..522..723C}, which allows for an interactive supervision and interpolation of the affected pixels.  Wavelength calibration, C- and S-distortion correction,
spectrum extraction, and atmospheric extinction correction were carried out using the \mbox{R\raisebox{-0.35ex}{E}D%
\hspace{-0.05em}\raisebox{0.85ex}{uc}\hspace{-0.90em}%
\raisebox{-.35ex}{{m}}\hspace{0.05em}E} package\footnote{See
\url{https://reduceme.readthedocs.io/en/latest/},
.}. Since the target was placed in different positions along the slit in each observing block, this allowed a first-order sky subtraction correction by subtracting the average image from each block. In addition, we employed the \texttt{xnirspec} tool \citep[][]{2003ApJ...584...76C} to perform a second-order sky subtraction by computing an averaged sky-spectrum of residuals which were subtracted by accounting for the specific image distortion at each pixel. 

Finally, the absolute flux calibration was obtained using the observation of the spectrophotometric standard star GD153. By fitting the upper boundary of the stellar continuum using the package \texttt{boundfit} \citep[][]{2009MNRAS.396..680C}, we derived not only the overall response curve, but also a normalized telluric absorption spectrum. A final flux correction factor of 1.32 was derived from the photometric calibration of the acquisition image using available Sloan $r'$ data. This factor accounts for light losses due to the seeing being slightly larger than the slit width.

\subsection{EMIR spectroscopic data}
The near-infrared observations with EMIR were carried out on 2020 June 18-19 (seeing~$\sim1.1$~arcsec),
using the K~grism, which provides and intermediate spectral resolution \mbox{$R\sim4000$}, with a 1.73~\AA/pixel sampling, and a spatial scale of 0.1945~arcsec/pixel. Two observing blocks consisting in three ABBA sequences/block with an exposure time of 360~sec per individual frame, accounted for a total exposure time of 8640~sec. The spatial separation between the~A and B~exposures were 7.4~arcsec. The initial data reduction was carried out using {\tt PyEmir} \citep[][]{2019ASPC..521..232P,2019ASPC..523..317C}, a dedicated pipeline specifically created for the reduction of EMIR data, including bad-pixel masking,
flat-fielding, image rectification and wavelength calibration. In addition to this, we also performed the absolute flux calibration, following the same procedure employed with the OSIRIS data. In this case the spectrophotometric standard star HD116405 was used. The target spectrum was accordingly corrected for telluric absorption, and a final flux correction factor of 1.11, to account for light losses in the slit, was computed using the acquisition image calibrated with available $K_{\rm s}$ 2MASS photometric data.

Both OSIRIS and EMIR spectra were corrected for Galactic extinction ($E(B-V)=0.015$) following \citet[][]{1999PASP..111...63F} and using $R_{V} = 3.1$.

\section{Results and Discussion}\label{sec:results}
\subsection{The First $z>3$ BL Lac Object}

The optical and NIR spectra of J1219 are shown in Figure~\ref{fig:osiris_emir}. For a comparison, we also overplot the SDSS spectrum. By comparing the wavelength dependent signal-to-noise ratio (SNR), one can notice the high SNR of the OSIRIS data which is $\sim$3 times better than the SDSS one. Since the EMIR spectrum is relatively noisy, a re-binning of 50 pixel was done to achieve a SNR of $\sim$5 per pixel at the center of the spectrum.

Both OSIRIS and EMIR spectra do not exhibit any emission lines suggesting the BL Lac nature of J1219 and more importantly {\it the first \gm-ray emitting BL Lac source identified beyond $z=3$}. Considering the beginning of the wavelength drop at $\sim$5552 \AA~as the Lyman-$\alpha$ break, the source redshift can be constrained as $z\sim3.57$. However, the onset is not only set by the redshift but also by proximity effects in the neutral gas distribution and by the spectral resolution. One would expect a large proximity zone of many Mpc of ionized gas around the quasar (due to the quasar), so that the Lyman-$\alpha$ forest will appear to start at a lower redshift than the quasar itself \citep[cf.][]{2008A&A...491..465D}. Therefore, the actual source redshift could be somewhere between 3.5$-$3.6. Based on the visual inspection of the SDSS spectrum, \citet[][]{2020ApJS..249....3A} reported the redshift of J1219 as $z=3.59$ and the same we have adopted in our analysis. 

We highlight the expected location of a few prominent broad emission lines, e.g., C~{\sc iv}, in Figure~\ref{fig:osiris_emir} though none of them are detected. One can notice a narrow emission line very close to the expected position of Lyman-$\alpha$ emission. This feature, however, is the residual of the bright [OI] sky line, which was subtracted during the OSIRIS data reduction. Furthermore, on a comparison with the SDSS spectrum taken on 2011 May 28, the continuum flux level was found to be lower by a factor of $\sim$3 during GTC observations. This finding suggests the absence of emission lines even during the low jet activity. This, in turn, hints a radiatively inefficient accretion process and hence strengthens the BL Lac nature of J1219.

We used the OSIRIS spectrum to determine the upper limit on the EW of the C~{\sc iv} emission line. The spectrum was brought to the rest-frame and fitted with a power-law to reproduce the continuum in the wavelength range [1500,1600] \AA. Additionally, the C~{\sc iv} emission line was modeled as a Gaussian function with a variable EW while keeping FWHM fixed to 4000 km s$^{-1}$, a value typical for blazars \citep[e.g.,][]{2012ApJ...748...49S,2016AJ....151...35L}. Then, we performed a $\chi^2$ test by varying EW and derived the upper limit to it when $\chi^2>\chi^2$ (99.7\%), i.e., at 3$\sigma$ confidence level. This exercise resulted in the EW upper limit of 3.4 \AA~(Figure~\ref{fig:osiris_emir}, bottom panel). We also computed the minimum detectable EW (EW$_{\rm min}$) following a model independent approach described in \citet[][]{2005AJ....129..559S}. In particular, the EW values were derived for wavelength intervals of fixed size (20 \AA) along the whole spectrum. The EW$_{\rm min}$ was considered as the 3$\sigma$ deviation from the mean ($\mu\sim0$ \AA) of the obtained distribution and found to be EW$_{\rm min}=1.9$ \AA. Altogether, any emission line, if exists, must have the rest-frame EW $\lesssim3.5$ \AA. We repeated the same exercise on the SDSS spectrum and estimated 3$\sigma$ EW upper limit and EW$_{\rm min}$ as $\sim$6 \AA~and 2.5 \AA, respectively. On the other hand, the results acquired from the EMIR data analysis are less constraining, due to low SNR, giving 3$\sigma$ EW upper limit and EW$_{\rm min}$ for H$_{\beta}$ line as $\sim$33 \AA~and 9 \AA, respectively.

\subsection{Intervening Absorption Systems}

There are three absorption features in the rest-UV continuum longward of the Lyman-$\alpha$ forest at $\sim$5770 \AA, $\sim$5870 \AA, and $\sim$6065 \AA~(Figure~\ref{fig:osiris_emir}). The most prominent absorption feature at $\sim$6065 \AA~is clearly the \ion{C}{4} $\lambda \lambda$1548,1551 doublet of a foreground absorber at $z=2.9135$, as indicated by the separation and relative strength of the lines as well as a clear Lyman-$\alpha$ absorption line at the same redshift. The other two absorption features are also possibly \ion{C}{4} doublet absorption at $z=2.7243$ and $z=2.7900$. However, the lower depth of the absorption lines, combined with noisy corresponding Lyman-$\alpha$ absorption in the Lyman-$\alpha$ forest makes it difficult to confirm. However, these two absorption features at $\sim5770$ \AA\ $\sim5870$ \AA\ do not correspond to known absorption lines at the redshift of J1219 that are commonly seen in the rest-UV of galaxies or quasars. The feature at $\sim5770$ \AA\ could be explained by \ion{Si}{2}$\lambda$1260, but there are no other apparent absorption features of low ions (\ion{O}{1}, \ion{C}{2}, \ion{Fe}{2} and other less prominent lines of \ion{Si}{2}). Thus, we deduce that there is no conclusive evidence of any absorption features associated with the blazar itself. 

We also note the strong signal at 3700$-$4176 \AA, below the Lyman limit of J1219. This indicates an intergalactic medium (IGM) transparent to Hydrogen-ionizing photons for at least 68 Mpc $h^{-1}$ (proper), nearly $1.5\times$ the measured mean free path of $\sim46$ Mpc $h^{-1}$ at this redshift \citep{2009ApJ...705L.113P}. This is also much larger than the typical ``proximity zone'' of ionized gas found around luminous quasars at similar redshifts \citep{2008A&A...491..465D}.

\subsection{Spectral Energy Distribution}
\begin{figure*}[t!]
\hbox{
\includegraphics[scale=0.41]{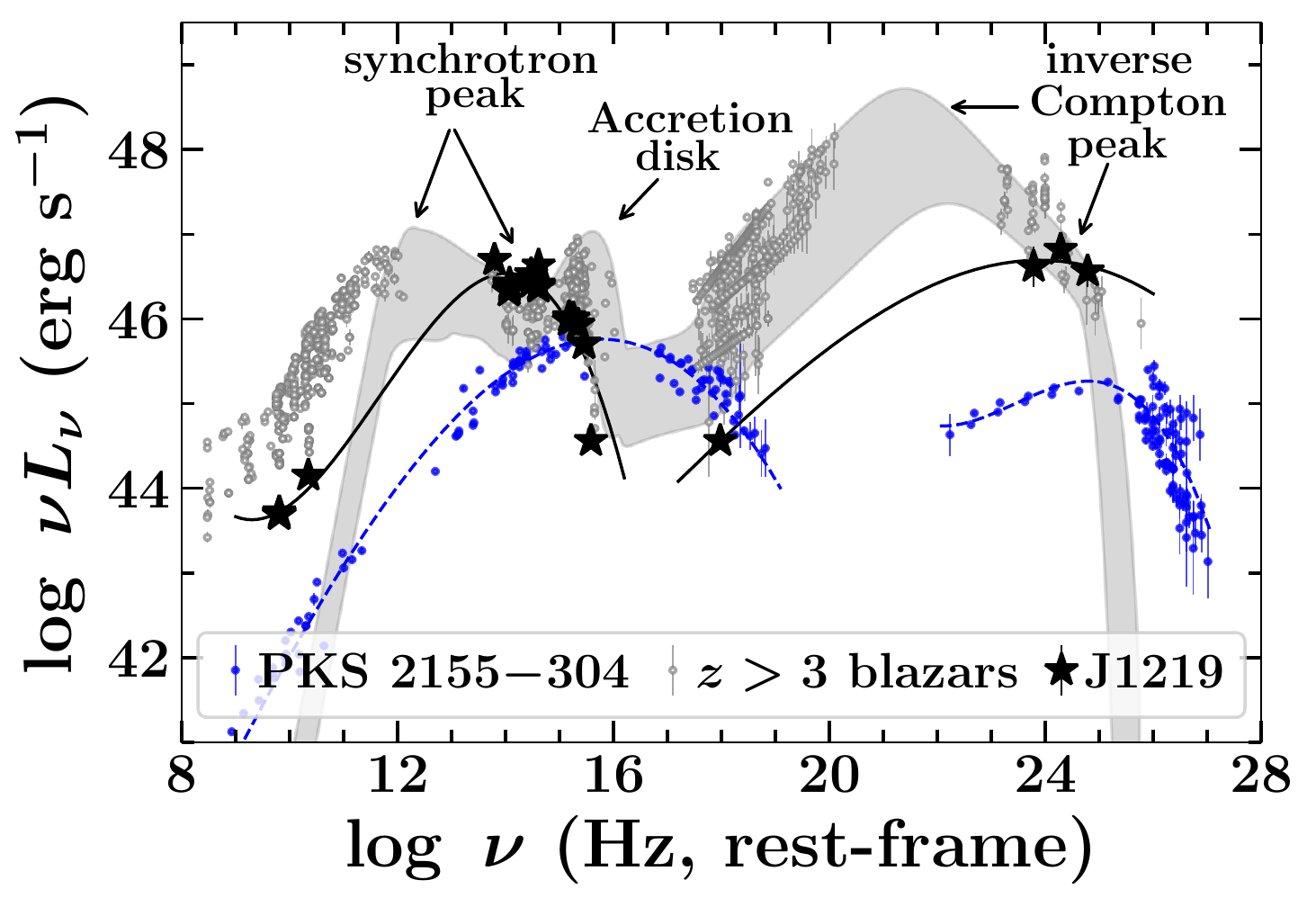} 
\includegraphics[scale=0.28]{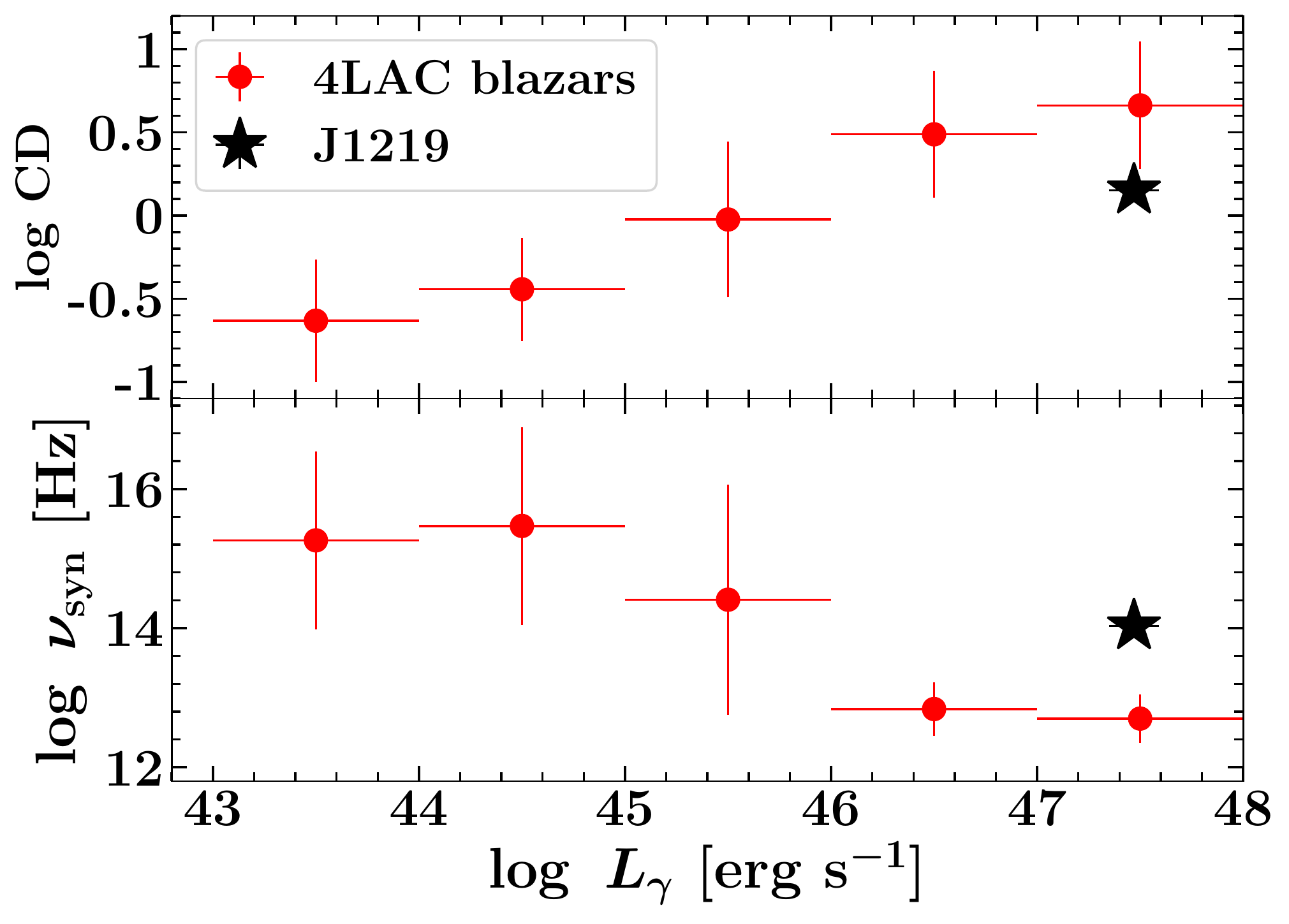} 
\includegraphics[scale=0.34]{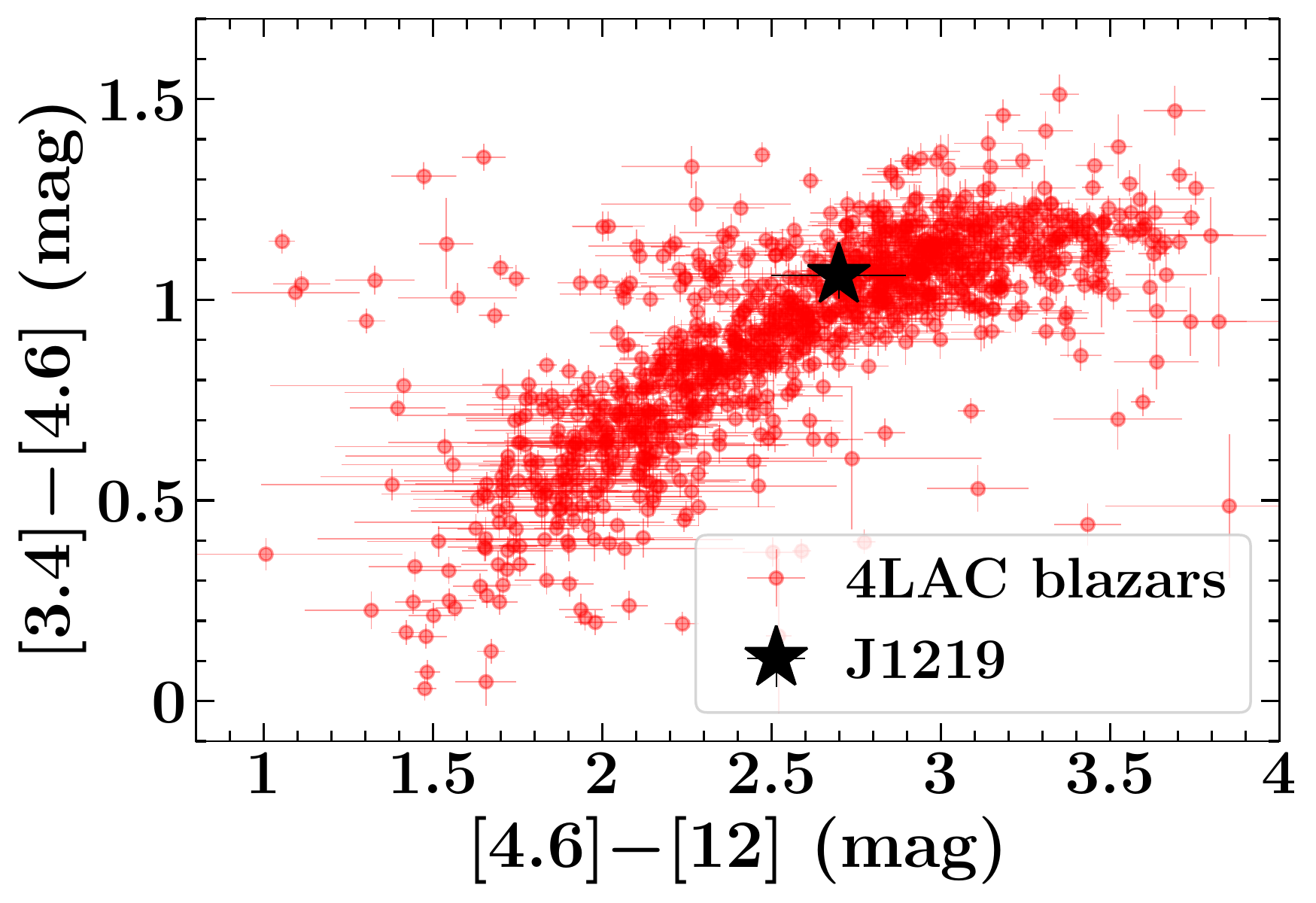} 
}
\caption{Left: the broadband SED of J1219 (black stars), 9 \gm-ray detected, $z>3$ FSRQs studied in \citet[][grey circles]{2020ApJ...897..177P}, and well-known BL Lac object PKS 2155$-$304 (blue circles). The fitted third-order polynomial to J1219 and PKS 2155$-$304 SEDs are shown with black solid and blue dashed lines, respectively. The shaded area represents the range of modeled SEDs of $z>3$ FSRQs to highlight the SED peaks. Middle: The location of J1219 in the updated \fermi~blazar sequence. Right: WISE color-color diagram of \fermi~blazars (red circles). The position of J1219 is shown with a black star. 
}
\label{fig:SED}
\end{figure*}

The broadband spectral energy distribution (SED) of J1219 using all available observations is shown in the left panel of Figure~\ref{fig:SED}. For a comparison, we also plot the SEDs of 9 other \gm-ray detected, $z>3$ FSRQs studied in \citet[][]{2020ApJ...897..177P}. The SED peak locations for J1219 were determined by fitting a third-order polynomial. For other sources, we used the results of the leptonic radiative modeling done in \citet[][]{2020ApJ...897..177P}. As can be seen, J1219 has both SED humps peaked at considerably higher energies compared to $z>3$ FSRQs. At optical-UV frequencies, a prominent thermal bump from the accretion disk is observed from other $z>3$ sources. The radiation from J1219, on the other hand, appears to be synchrotron dominated with no traces of the accretion disk emission. Furthermore, the Compton dominance, which is the ratio of the inverse Compton to synchrotron peak luminosities, of $z>3$ FSRQs is significantly larger ($\sim$100) than that is observed for J1219 ($\sim 1$).

Since emission lines are not detected in the optical/NIR spectra of J1219, we have derived the 3$\sigma$ upper limit on the C~{\sc iv} line luminosity following the same procedure adopted to compute the EW upper limit and found it to be $\sim2.5\times10^{42}$ \lum~($\sim1.1\times10^{43}$ \lum, when using SDSS data). Considering the line flux ratios from \citet[][]{1991ApJ...373..465F} and assuming a 10\% BLR covering factor, the 3$\sigma$ upper limit on the accretion disk luminosity is $\sim2.2\times10^{44}$ \lum. Assuming the central black hole mass as 10$^9$ \Msun~\citep[][]{2020ApJ...897..177P}, we get $\dot{L}_{\rm acc}/L_{\rm Edd}<0.002$ or 0.2\% of $L_{\rm Edd}$. A relatively high upper limt of 2\% of $L_{\rm Edd}$ was obtained when $H_{\beta}$ line luminosity derived from the EMIR data ($\sim10^{43}$ \lum) was considered. Such a low level of radiatively inefficient accretion is unlikely to photo-ionize the BLR clouds and can explain the observed lack of emission lines. 

We comment on the location of J1219 in the blazar sequence by comparing its SED with the BL Lac PKS 2155$-$304 ($z=0.12$) and other $z>3$ blazars (Figure~\ref{fig:SED}, left panel). PKS 2155$-$304 exhibits a low-luminous and high-frequency peaked SED compared to that of J1219. Other $z>3$ FSRQs are more powerful and their SED peaks are located at even lower energies, as expected from the blazar sequence. Furthermore, since the updated \fermi~blazar sequence is based on the sources grouped in different \gm-ray luminosity ($L_{\gamma}$) bins \citep[][]{2017MNRAS.469..255G}, we have calculated synchrotron peak frequency and Compton dominance for all blazars present in the 4LAC catalog and grouped them in different $L_{\gamma}$ bins (Figure~\ref{fig:SED}, middle panel). Interestingly, compared to other luminous blazars ($L_{\gamma}>10^{47}$ \lum), J1219 has a relatively high-frequency peaked and a low-Compton dominated SED. The physical behavior of J1219, therefore, deviates from the predictions of the \fermi~blazar sequence. This stresses the importance of searching high-redshift BL Lacs to critically examine the proposed sequence.

The \gm-ray emitting blazars are known to occupy a distinct place in the Wide-field Infrared Survey Explorer (WISE) color-color diagram \citep[e.g.,][]{2011ApJ...740L..48M}. In the right panel of Figure~\ref{fig:SED}, we show this plot for 4LAC blazars and show the location of J1219. As can be seen, J1219 is located right in the middle of the WISE blazar strip indicating a close connection between the IR and \gm-ray emissions of the source.

\subsection{Number density of BL Lacs beyond $z=3$}
The space density of $\gamma$-ray emitting BL Lacs has only been constrained to $z\approx$2.5 and measured there to be $\approx$0.01\,Gpc$^{-3}$ \citep{2014ApJ...780...73A}. Extrapolating the proposed evolution to $z=3.5$ finds a space density of an order of magnitude lower, $\approx 0.001$\,Gpc$^{-3}$. Considering the comoving volume within that redshift, this implies a prediction that there should be between one and two BL Lacs in the entire Universe at that redshift, and this work has discovered one of them. Therefore, identification of more of $z>3$ \gm-ray emitting BL Lac sources will challenge our current understanding about their cosmic evolution.

\acknowledgments
We are thankful to the referee for a constructive criticism. VSP's work was supported by the Initiative and Networking Fund of the Helmholtz Association. A.D. acknowledges the support of the Ram{\'o}n y Cajal program from the Spanish MINECO. CC, NC JG, and AGdP acknowledge financial support from the Spanish Programa Estatal de I+D+i Orientada a los Retos de la Sociedad under grant RTI2018-096188-B-I00, which is partly funded by the European Regional Development Fund (ERDF). We are grateful to IAC Director for approving our DDT request. This work is based on observations made with the GTC telescope, in the Spanish Observatorio del Roque de los Muchachos of the Instituto de Astrofísica de Canarias, under Director's Discretionary Time.

\bibliographystyle{aasjournal}

\end{document}